# Universal and dynamic ridge filter for pencil beam scanning particle therapy: a novel concept for ultra-fast treatment delivery

Vivek Maradia[1,2], Isabella Colizzi[1,2], David Meer[1], Damien Charles Weber[1,3,4], Antony John Lomax[1,2], Oxana Actis[1], Serena Psoroulas[1]

[1]Paul Scherrer Institute, Switzerland
[2]ETH Zurich, Switzerland
[3]University Hospital Zurich, Switzerland
[4]University Hospital Bern, University of Bern, Switzerland

Vivek Maradia and Isabella Colizzi shared first authorship

## Abstract

**Purpose**

In PBS particle therapy, a short treatment delivery time is paramount for the efficient treatment of moving targets with motion mitigation techniques (such as breath-hold, rescanning, and gating). Energy and spot position change time are limiting factors in reducing treatment time. In this study, we designed a universal and dynamic energy modulator (ridge filter, RF) to broaden the Bragg peak, to reduce the number of energies and spots required to cover the target volume, thus lowering the treatment time.

**Methods**

Our RF unit comprises two identical RFs placed just before the isocenter. Both RFs move relative to each other, changing the Bragg peak's characteristics dynamically. We simulated different Bragg peak shapes with the RF in TOPAS and validated them experimentally. We then delivered single-field plans with 1Gy/fraction to different geometrical targets in water, to measure the dose delivery time using the RF and compare it with the clinical settings.

**Results**

Aligning the RFs in different positions produces different broadening in the Bragg peak; we achieved a maximum broadening of 2.5 cm. With RF we reduced the number of energies in a field by more than 60%, and the dose delivery time by 50%, for all geometrical targets investigated, without compromising the dose distribution transverse and distal fall-off.

**Conclusions**

Our novel universal and dynamic RF allows for the adaptation of the Bragg peak broadening for a spot and/or energy layer based on the requirement of dose shaping in the target volume. It significantly reduces the number of energy layers and spots to cover the target volume, and thus the treatment time. This RF design is ideal for ultra-fast treatment delivery within a single breath-hold (5-10 sec), efficient delivery of motion mitigation techniques, and small animal irradiation with ultra-high dose rates (FLASH).

## Introduction

Particle therapy uses the Bragg peak and thus their more localized dose deposition in tissues, thus delivering more conformal radiation delivery and less integral dose. Pencil beam scanning (PBS) is the most advanced particle therapy delivery technique which was brought into clinical practice at Paul Scherrer Institute (PSI) in 1996 [1], [2]. PBS provides better dose distribution conformity and also better spares healthy tissues when compared with equivalent photon plans [3], [4]. However, PBS is especially prone to uncertainties caused by anatomical changes and target/tissue motion during the treatment [5]–[7]. Guidelines for treatments of moving target explicitly require motion mitigation techniques to reduce the interplay effect between the motions of the patient and the beam delivery [8].

PBS treatment delivery could benefit considerably from shorter treatment delivery times. A shorter field delivery duration will improve patient comfort. Hypofractionation with short treatment delivery time could also increase patient throughput and reduce treatment costs for a relatively expensive modality [9]. Shorter treatment delivery times would increase efficiency of treatments using motion mitigation techniques such as gating [10], rescanning [11]–[13], and breath-hold [14], [15] or any combination of these techniques.

Treatment delivery time with PBS depends both on the beam-on time and the dead time (the time required to change energy layers and/or lateral position) between pencil beams. Beam-on time can be reduced significantly by using high-intensity beams [16]–[19] and dead time for lateral position change can be reduced by reducing the number of spots, e.g. using spot-reduced treatment planning [20]. Combining both could even enable single (eventually hypofractionated) fields to be delivered within a single breath-hold [21], and could be an important enabler of proton FLASH techniques [22], [23]. However, with shorter treatment times the time required to change energy becomes a limiting factor.

Particle therapy centers use cyclotrons, with fixed extraction energy, or synchrotrons, which can adjust energy pulse-to-pulse. A synchrotron-based facility produces sharper Bragg peaks at lower energies compared to a cyclotron-based facility same holds for particle beams (e.g. carbon and helium) vs proton beams, as well as low energies vs higher energies [24]. This will have an impact on the number of energy layers required for planning, and, in turn, on the irradiation time. For cyclotron-based facilities, the energy change time ranges from 80 ms to 900 ms [25]. However, for most synchrotron-based facilities, it is even slower ranging from 200 ms to 2000 ms [25]. Thus, reducing the number of energy layers would improve the beam delivery efficiency.

One way to reduce the number of energy layers (and in turn, the number of spots) required to cover the full target volume is to broaden the Bragg peak by using a ridge filter (RF). Over the last two decades, many RF designs were proposed and used clinically for particle therapy [26], some field- and patient-specific [27]–[29], others more universal [30]–[32]. Moreover, in the last couple of years interest in RFs increased significantly due to growing interest in ultra-high dose rate

(FLASH) experiments with animals and for planning studies with real patient cases [33], [34]. The drawback of RFs is increased lateral and distal fall-offs of the beam, which can affect dose conformity. Therefore, we propose a new universal RF design to dynamically modulate, within the delivery of individual fields, the characteristics of the Bragg peak (i.e. broadening, distal fall-off of the beam), for a spot and/or energy layer based on the requirement of the dose shaping.

In this article, we propose a new RF unit that comprises two identical RFs placed just before the isocenter. Both RFs are movable relative to each other to change the Bragg peak's characteristics dynamically. We designed the RF using Monte Carlo simulations validated against measurements performed at the PSI Gantry 2 facility. We then delivered single-field plans with and without RF to three different geometrical targets in water, to evaluate uniformity, penumbra, and treatment time reduction. In this article, we studied the performance of our dynamic RF only with protons. However, we expect the method described in this paper to be applicable to other particle therapy facilities as well.

## Methods

**Concept behind the new RF design:**

Our dynamic RF unit consists of a first (M) and a second (N) RF. These RFs are arranged transversally in line after each other along the proton beam path. The M-RF is movable relative to the N-RF to vary the energy/momentum spread of beam dynamically. Depending on the relative positions of the RFs, protons will see different thicknesses of material, and therefore loose different energy and generate different Bragg peaks with variable broadening. Figure 1 illustrates a proposed RF unit comprising M- and N-RF.

In Figure 1(a), the RFs M and N are displaced relative to each other in such a way as to cause minimum broadening in the Bragg peak, by positioning the M-RF so that the peaks of its ridges align with the valleys of the N-RF. In this case, protons passing through the M-RF at a peak will pass through the N-RF in a valley and vice versa. In this configuration, the RF unit acts as a range shifter.

In Figure 1(b), the RFs M and N are displaced relative to each other in such a way as to cause maximum broadening in the Bragg peak, by positioning the M-RF so that the peaks of its ridges align with the peaks of the N-RF. When matching peak with peak and valley with valley, we will get maximum energy/momentum spread.

The displacement between the RFs can be varied continuously between the extremes shown in Figure 1(a) and Figure 1(b), respectively; thereby moving the two ridges relative to each other, any possible broadening between minimum and maximum can be obtained, as desired for the treatment.

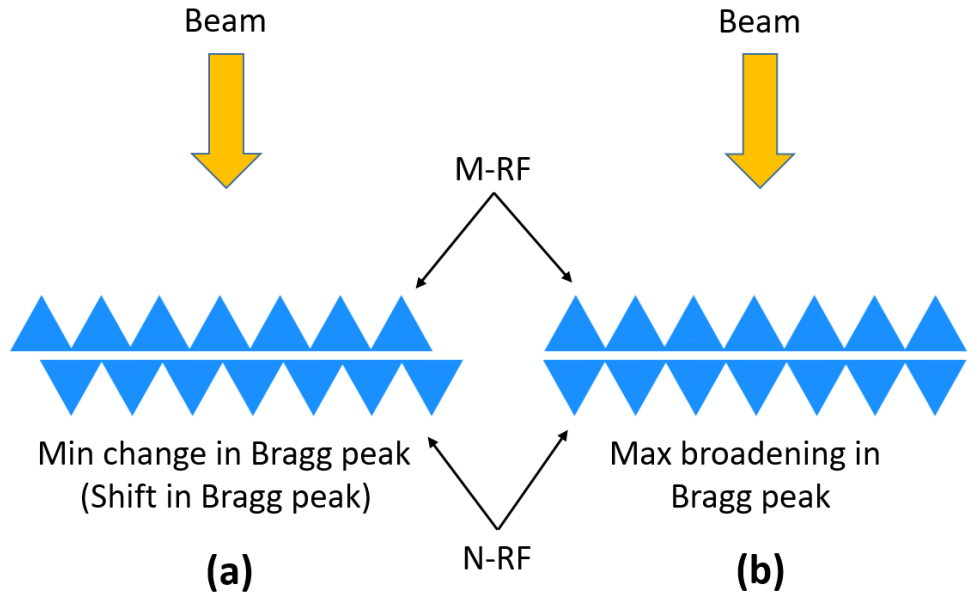

*Figure 1 Different configuration of RF.*

**Realization**

The RF shape was created through an iterative design process with the aim of achieving a uniform depth dose curve spread of a several millimeters with a fall-off comparable to a pristine Bragg peak for all energies commonly utilized for treatments at our clinic (70-230 MeV). Selecting the material was the first step in designing the RF. It was selected to reduce the multi-Coulomb scattering-related increase in beam size. The scattering probability decreases for material with high radiation length. Therefore, Polyethylene would have been the ideal material due to its low radiation length (44 g/cm2) and low density (0.88-0.96 g/cm3). However, it is incompatible with stereo lithography (SLA) 3D printing, so we utilized Resin (40 g/cm2 and 1.19 g/cm3). Second, we determined the period of the ridges and the maximum broadening we wished to obtain. The former is influenced by the material's water-equivalent depth, while the latter is influenced by the beam size. We were able to produce a depth dose curve independent of the beam position by selecting a maximum thickness of 4 cm, 2 cm per single ridge, and a period of 0.55 cm. Finally, the final dose distribution was produced by optimizing the RF shape (e.g. the base thickness and ridge slope), taking into account printing restrictions on the peak width.

Figure 2 shows one unit of the designed RF, which has a maximum thickness of 2 cm and minimum of 0.5 cm. It features a one-dimensional periodic structure with a 0.55 cm period orthogonal to the beam direction. The RF was realized utilizing stereo lithography (SLA), a 3D printing technique that results in higher resolution shapes than conventional 3D printing. In Figure 2, we present a picture of the printed design.

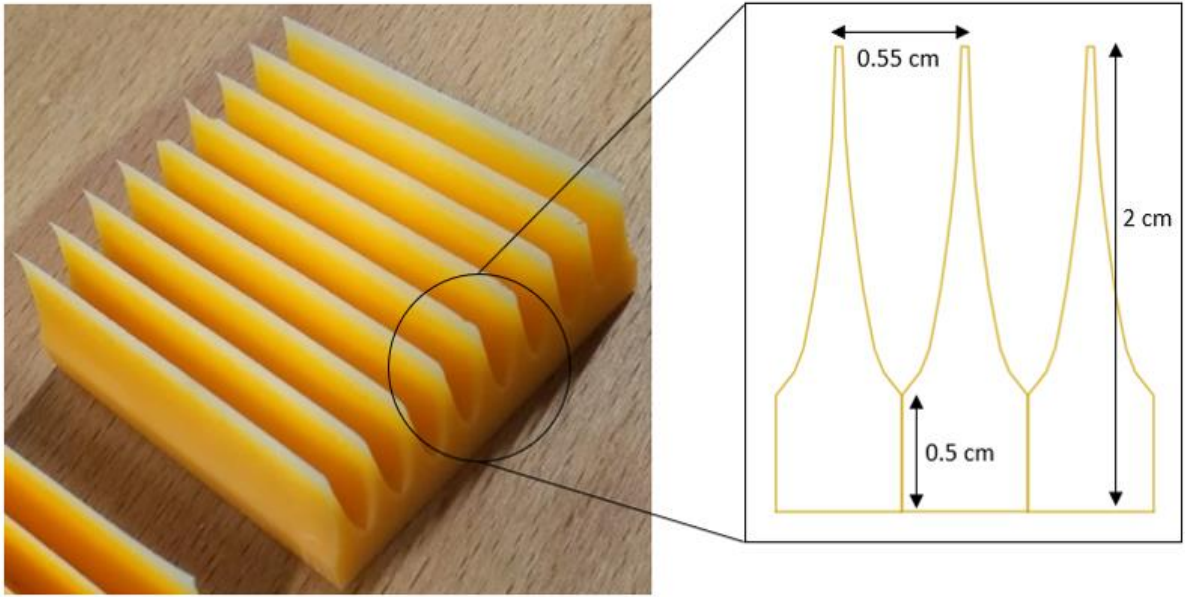

*Figure 2 RF printed with SLA technique and dimensions of the RF*

**Simulation environment**

In the design process, to generate the depth dose curves and the phase space of the beam with RF, taking into account multi-Coulomb scattering, secondary particle production, and energy straggling effects, we used Monte Carlo simulations performed in TOPAS (TOol for PArticle Simulation) [35], an extension for radio- and proton therapy of the Geant4 simulation toolkit [36]. We utilized the suggested [37] default modular physics list for the simulation given in this study (G4EMLOW6.48, G4NDL4.5, PhotonEvaporation3.2, RadioactiveDecay4.4, G4SAIDDATA1.1, G4NEUTRONXS1.4, G4PII1.3, G4ABLA3.0, G4ENSDFSTATE1.2.1, G4TENDL1.0). To decrease statistical error, simulations with 10 million protons were performed.

**Experimental setup**

Proton beam measurements of Bragg curves and lateral dose profiles were carried out at PSI Gantry 2 [38]. Depth dose curves in water were measured using a large plane-parallel ionization chamber (PTW 34070) mounted on a range scanner (Figure 3(a)), whilst a charge coupled device (CCD) camera with a scintillating screen was used to measure the lateral beam profiles. PMMA plates were installed on top of the scintillating screen to measure lateral profiles at different depths (Figure 3(b)). Nozzle to isocenter distance was maximized (47.6 cm) to increase the beam size before the RF and reduce the risk of ripples in the transverse dose distribution, whilst the RF was placed close to the isocenter (1 cm). This has a beneficial effects on the lateral penumbra but alternatively might cause ripples. However, we found that shifting the RF up to 10 cm away from the isocenter has virtually no effect on the beam size, therefore this choice does not affect our results.

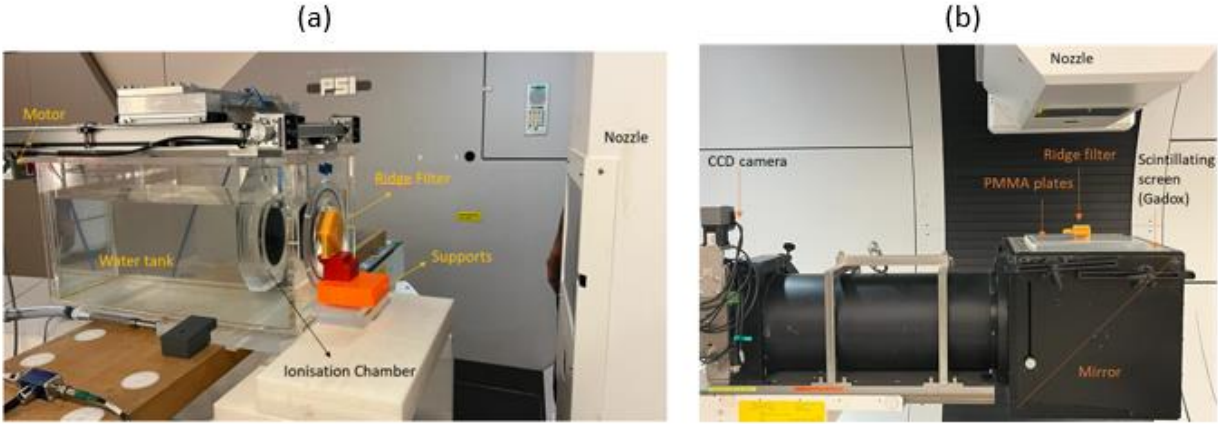

Figure 3 Experimental setup: (a) range scanner with RF placed closed to the isocenter, (b) CCD camera with PMMA plates and RF.

**Phantom Planning Study**

The spot positioning and the dose calculation on geometrical targets were carried out using an in-house developed MATLAB script for experimental use. Using this code, we were able to calculate the dose distribution in water for simple geometries placed at various depths. Each plan was first calculated using pristine Bragg peaks and then utilizing the new depth dose curves generated with the RF. The dose distribution was calculated on a regular spot grid (in the transverse plane) of 0.4 cm for a single field. In *Table 1*, we illustrate the target shape, size, position, and the delivered dose.

Table 1 Target's dimension (T and U are the dimensions perpendicular to the beam direction (z)), position and dose prescription

| Target | Dimensions (Volume, T x U x z) | Half-SOBP depth [cm] | Dose [Gy] |
|---|---|---|---|
| Sphere | 904 cm$^3$ (r = 6 cm) | 15 / 10 | 1 |
| Box 1 | 4080 cm$^3$ (12 cm x 20 cm x 17 cm) | 12.5 | 1 |
| Box 2 | 272 cm$^3$ (4 cm x 4 cm x 17 cm) | 12.5 | 1 |
| Box 3 | 64 cm$^3$ (4 cm x 4 cm x 4 cm) | 11/13/15/17 | 1 |
| Cylinder | 214 cm$^3$ (r = 2, l = 17 cm) | 12.5 | 1 |

**Calculation of the delivery time**

By studying the machines log files, we were able to determine the dose delivery time and to isolate all contributions to treatment time (spot changing time, energy layer changing time, and beam-on time). The beam-on time is the time required to deliver the dose and the dead time is the sum of spot and energy layers changing time. The time required to change the spot location in the U and T directions determines the spot changing time.

**Spread out Bragg peak (SOBP) generation**

Even though our RF allows all possible combinations, in this study we decided to stick to the two simplest cases: full maximum broadening, and a combination of maximum and minimum broadening. We generated SOBPs in these two configurations and evaluated their quality (against no-RF SOBPs) looking at two quantities: first, in terms of dose uniformity [39], defined as the ratio:

$$DU = \frac{d_{max} - d_{min}}{d_{max} + d_{min}} \cdot 100 \tag{1}$$

where $d_{max}$ and $d_{min}$ are the maximum and minimum dose within the SOBP width (defined as the distance between the proximal and distal 90% dose) [39], respectively; second, looking at distal fall-off, defined as the distance along the beam axis where the dose in water reduces from 80% to 20%. The number of required energies varies depending on the desired uniformity. However, increasing the number of energies lengthens the treatment delivery time. Consequently, a good compromise between time reduction and dosage homogeneity is required. In our case, we opted to reduce the number of layers while keeping a dose uniformity below 1%.

## Results

In this section, we will discuss the findings of our simulations and compare them to measurements. We will start by examining the depth dose curves generated with three distinct RF's configuration, then look at the SOBP in terms of lateral fall-off and uniformity. The influence of the RF on the dose profile will be investigated, and the penumbra and flatness will be compared to dose distributions without RF. Following that, we will demonstrate experimentally how, using the RF, we may reduce the delivery time for geometrical phantoms.

**Effect of RF on the integral depth-dose curves**

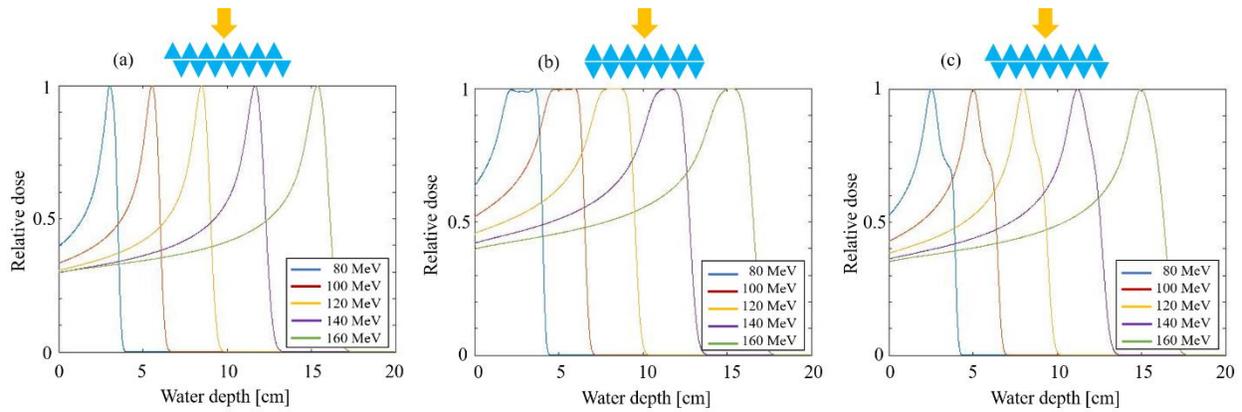

*Figure 4 TOPAS simulations of the depth dose curve with the RFs aligned in three different positions.*

Depending on the position of RFs we can generate different kinds of broadening in the depth dose curve. In Figure 4, we show the simulated Bragg peak in water after the beam has gone through three distinct RF's configurations. In Figure 4(a), the M-RF valleys are aligned with the N-RF peaks, resulting in negligible Bragg peak broadening. The Bragg peak width (defined as the distance between the proximal and distal 80% dose curves) [39] is on average 0.9 cm. In Figure 4(b) the M-and N-RFs are aligned, resulting in maximum depth dose curve broadening (with an average width of 2.65 cm). One RF-generated depth dose curve can replace up to ten pristine Bragg curves. Figure 4(c) shows an intermediate position shift of the M-RF peaks with respect to the N-RF peaks, which results in a widened Bragg curve. Due to the current experimental limitations of not having a motorized drive to accurately align the two RFs, we have benchmarked only the minimum and maximum broadening configurations with data.

The measured depth dose curves produced with the RF in the position of minimal broadening are compared against TOPAS simulations in Figure 5(a). The measured distal fall-off confirms expectations at low energies, but it is smaller at high energies. The width of the peak is unchanged. However, for the configuration where the peaks of one RF match the peaks of the other, we find a small difference between the measured and generated curves (see Figure 5(b)). The distal fall-off is consistent with simulations, although the observed curve uniformity varies from 3-4 % for low-energy beams to 1 % for high-energy beams.

We assume that different momentum spread values cause the mismatch in the distal fall-off between measurement and simulations. The deviation in uniformity is probably due to the limited quality of the printing.

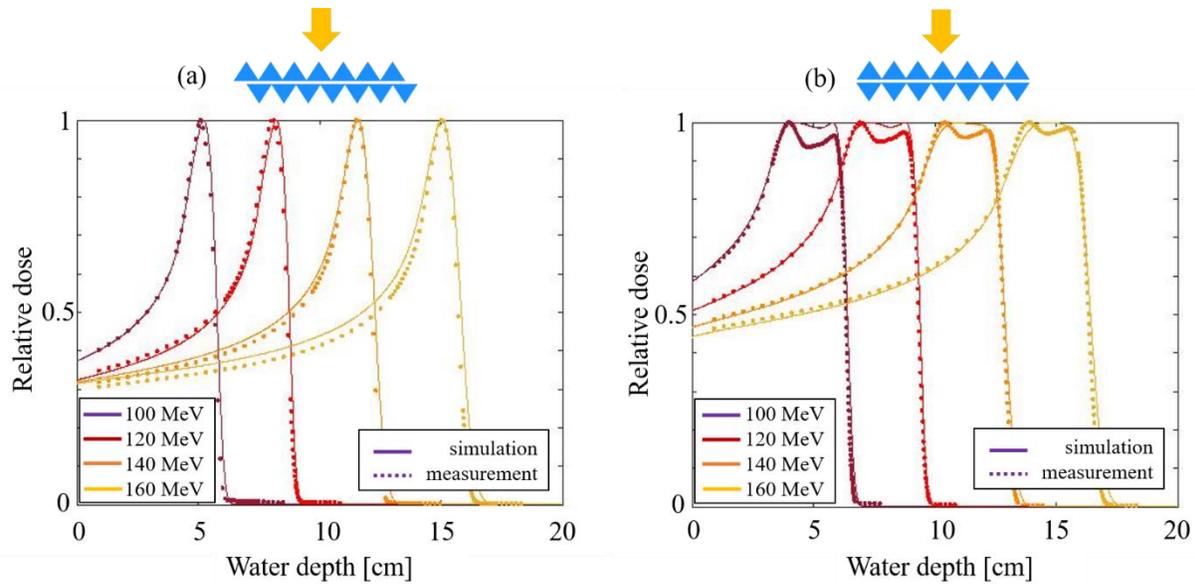

*Figure 5 Comparison between measurements (dot line) and TOPAS simulations (full line). The RF is positioned in the minimum (a) and maximum (b) broadening configuration.*

**SOBP comparison**

Figure 6 shows a comparison between a 12 cm SOBP generated with ((b) and (c)) and without RF (a). Additionally, the SOBP with the RF ((b) and (c)) are compared with measurement and show a good agreement with the simulations results, except for a small mismatch of distal fall-off, as previously discussed.

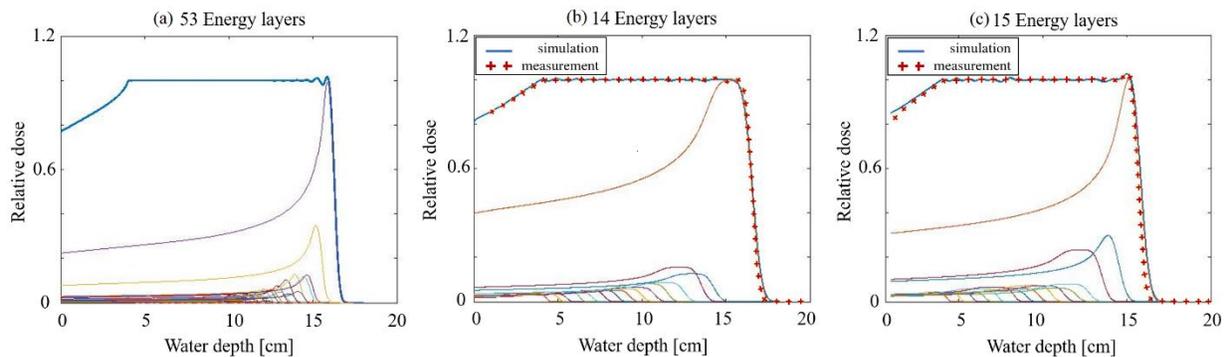

*Figure 6 12 cm (4-16 cm) SOBP generated with and without RF is compared. (a) SOBP created without RF with 53 energy layers (1.5 MeV energy steps). (b) SOBP generated with the RF in the configuration of the M- and N-RFs being aligned. (c) SOBP achieved by combining the maximum and minimum broadening depth dose curves. In (b) and (c) the simulated curve is benchmarked with the measured data.*

Without RF, 53 energy layers (with 1.5 MeV energy steps) are required to achieve a DU better than 1%. This is due to the high number of low weighted pristine Bragg peaks in the center of the target. With the help of the RF, we can lower the energy layers required to 15, while maintaining a uniformity below 1%. In this configuration, though, the distal fall-off increases from 0.56 cm to 0.72 cm, as shown in Figure 6(b). However, the dynamic nature of our RF allows combining two configurations (maximum and minimum broadening); in this third scenario (a sort of 'hybrid' configuration) we can limit the increase in distal fall-off and achieve a result similar to that obtained without RF (0.6 cm), as illustrated in Figure 6(c). On the other hand, this will increase the number of energy layers by just one or two in comparison to scenario (b).

The comparison data-simulation has been performed for several SOBP at different depths. We found, that with the help of the RF, we can lower the number of energy layers from 63 % to 75 %, depending on the dimension and position of the target, without compromises in distal fall-off and uniformity of the SOBP.

**Effect of the RF on the Lateral fall-off and flatness**

The dose profile perpendicular to the beam direction was measured with the CCD camera at various depths of a cubic and a spherical target in a Plexiglas phantom irradiated without and with the RF (in the maximum broadening configuration). Figure 7 shows the dose distribution with RF as well as the horizontal and vertical profiles. We compare the two distributions in terms of penumbra (the distance perpendicular to the beam axis at which the dose in water at a certain depth drops from 80% to 20%) [39] and flatness (defined as:

$$F_L \equiv \left( \frac{d_{max} - d_{min}}{d_{max} + d_{min}} \right) \times 100 \qquad (2)$$

with $d_{max}$ and $d_{min}$ the maximal and minimal dose about the center of the beam profile over 80% of the FWHM at a depth of interest, respectively) [39].

The sphere and box measured dose profiles (Figure 7(a) and (b)) are comparable to the profiles without RF. For both the x- and y-profiles, the flatness is less than 1%, and the penumbra is nearly unchanged with respect to the case without RF.

In particular, for the sphere, the penumbra is 0.63 cm for the horizontal profile and 0.65 cm for the vertical profile. The horizontal profile is the same with RF as without, but the vertical profile is 1.04 times larger due to the ridges of the RF. For the box, the penumbra is 0.57 in the horizontal and 0.6 cm in the vertical profile with and without RF.

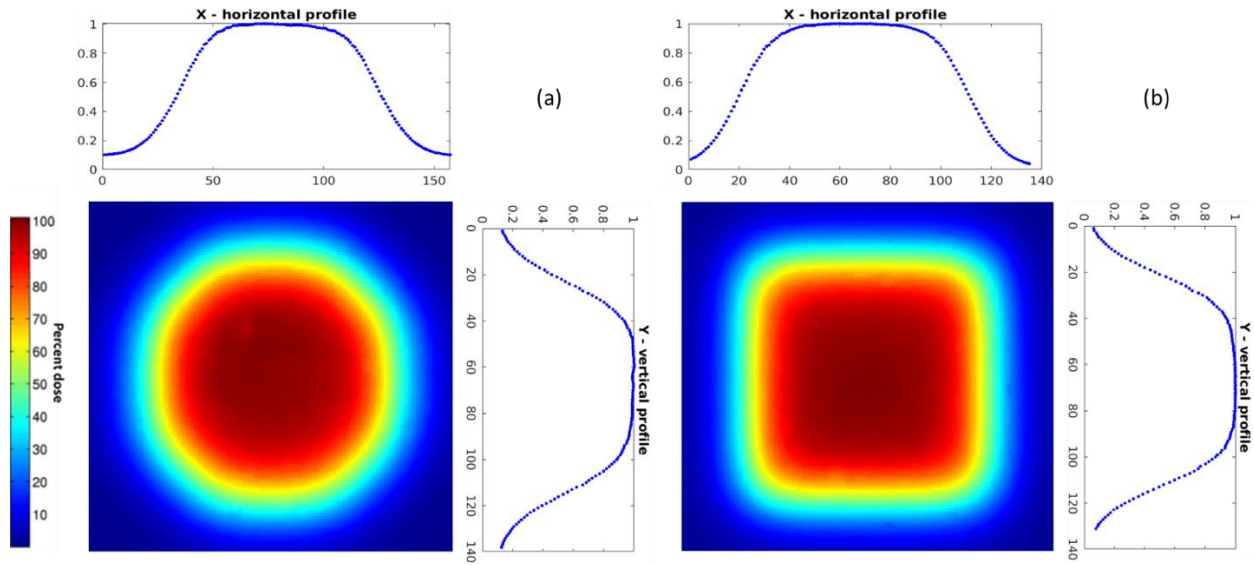

*Figure 7 Dose profile perpendicular to the beam direction with the RF. (a) Sphere of 2 cm radius measured at 17 cm depth. (b) Box of 4 cm lateral size measured at 16 cm depth.*

**Delivery time reduction with RF**

To prove experimentally the advantages of the use of the RF, we deliver different plans created with the geometrical shapes presented in Table 1 and measured the delivery time with and without RF.

Table 2 shows how the RF may be utilized to reduce dead time and examines its impact dependent on the form, size, and depth of the target. In Figure 7, we compare the total delivery time analyzing each contribution (beam on time, spot and energy layer changing time) with and without RF separately. In this work, we consider only the configuration of maximum broadening; however, since the 'hybrid' configuration of Figure 6(c) increases the number of energy layers only by 1 or 2, these delivery times are basically valid also for the hybrid configuration.

Table 2 Comparison between the field characteristics for plans on simple geometrical targets in water with and without RF: number of spots, number of energy layers, dead time (given by the sum of spot and energy layers changing time), and delivery time decrease

| Object (depth) | Number of spots without RF | Number of spots with RF | Number of energy layers without RF | Number of energy layers with RF | Dead time decrease [%] | Delivery time decrease [%] |
|---|---|---|---|---|---|---|
| Sphere (15 cm) | 14049 | 5425 | 30 | 24 | 54.8 | 40.1 |
| Sphere (10 cm) | 14017 | 5220 | 30 | 24 | 56.1 | 45.6 |
| Cylinder | 2898 | 1104 | 29 | 16 | 56.4 | 44.2 |
| Box 1 | 56982 | 22736 | 42 | 16 | 61.5 | 42.1 |
| Box 2 | 3402 | 1296 | 42 | 16 | 57.6 | 44.4 |
| Box 3 (17 cm) | 810 | 243 | 10 | 3 | 65.5 | 50.1 |
| Box 3 (15 cm) | 810 | 243 | 10 | 3 | 64.7 | 50.1 |
| Box 3 (13 cm) | 810 | 243 | 10 | 3 | 65.2 | 50.8 |
| Box 3 (11 cm) | 810 | 243 | 10 | 3 | 64.8 | 50.7 |

Compared to a plan without a RF, a plan with a RF reduces the number of energy layers and, as a result, the number of spots. Depending on the shape and position of the target, the number of spots can be lowered from 60 % to 70 %, and the energy layers from 17 % to 70 %. The overall dead time reduction for boxes is slightly higher than for spherical targets, but it is more than 50 % in both cases (it varies from 55 % to 65 %). This is because the most significant benefit for the spherical target is a reduction in the number of spots (up to 62%) while the number of energy layers required remains almost unchanged. For box targets, on the other hand, energy layer time reductions contribute almost as much as spot number reduction, thus resulting in a higher dead time reduction.

Overall delivery time can be reduced by 40-50 % for all evaluated targets; however, the reduction is greater for boxes and more superficial targets. With the help of the RF, the delivery time for geometrical targets of 200 cm$^3$ or less can be reduced to less than 10 seconds, as shown in Figure 8.

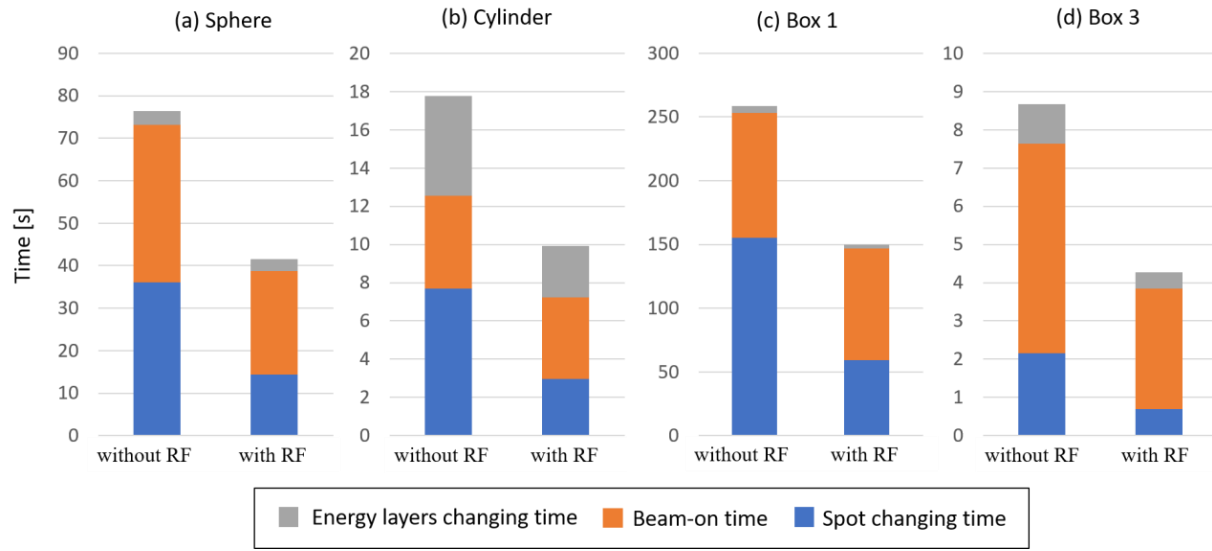

*Figure 8 Comparison between the delivery time with and without RF for three simple geometrical water phantoms: (a) sphere of 904 cm³ at 10 cm, (b) Cylinder of 214 cm³ at 12.5 cm, (c) Box of 4080 cm³ at 12.5 cm, (d) Box of 64 cm³ at 11 cm.*

**Robustness analysis**

- **Beam angle**

In Gantry 2, since the steering magnets are situated upstream of the last bending magnet, the beam is almost completely perpendicular to the patient (parallel scanning), with only small variations of a few mrad. We investigated how the Bragg curve changes when the beam has an angle of 2 or 4 mrad, which is the maximal variation in Gantry 2 PSI. The results for a 100 MeV beam are shown in Figure 9(a). We found that this divergence has no effect on the depth dose curve. Furthermore, we found that beam angles up to 35 mrad do not affect the depth dose curve. Therefore, our RF could be used in gantries with downstream scanning for small targets.

- **Spot scanning**

As the energy of the beam diminishes, the beam's size increases. Due to the 0.55 cm width of a single ridge, we may see differences depending on whether the beam's center is in the valley or the peak. We investigate this phenomenon by simulating the depth dose curve in water for different energies. As example, we illustrate in Figure 9(b) how the location of the beam's center affects the depth dose curve's shape of a 100 MeV beam ($\sigma_x \approx \sigma_y \approx 0.48$ cm in air at the isocenter). As shown, we may move the beam in the U and T directions without compromising the uniformity of the SOBP, and thus we can change the spot position without affecting the depth dose curve.

- **Misalignment uncertainty**

The effect of the misalignment between the two sides of the RF on the shape of the SOBP is an important error to consider. We simulated the effect of shifting the upper side by 0.5 mm and 1 mm. The findings for a 100 MeV beam are shown in Figure 9(c). We observe that a 1 mm misalignment has a non-negligible impact on the Bragg curve's shape. The initial dose is lower, and homogeneity is reduced to +/- 10%. For lower energies, the effect is more pronounced. To avoid RF alignment issues, the system should be able to set the ridges position with a precision better than 0.5 mm.

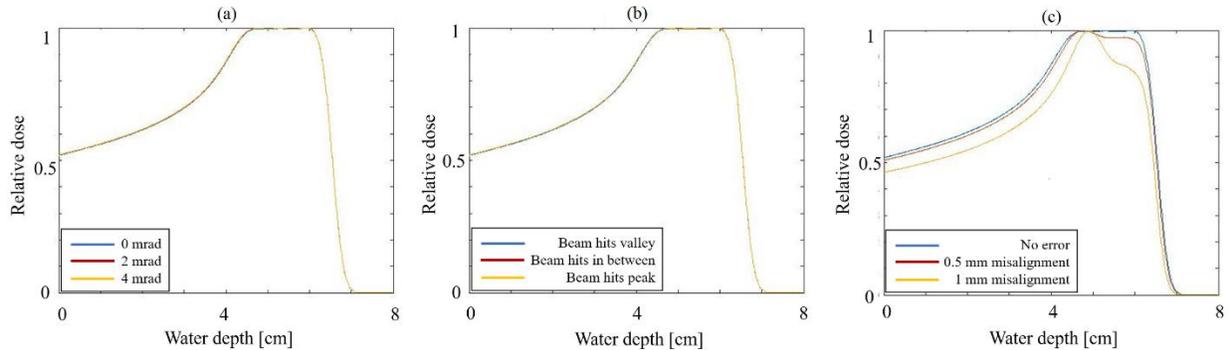

*Figure 9 Analysis of the robustness of the RF under possible errors and misalignments for a 100 MeV beam. (a) The beam hits the RF with 0, 2 and 4 mrad angle to the normal of the RF's surface. (b) The beam's center is on the valley, on the peak or on a position between the two. (c) The M-RF is not aligned with the N-RF, we consider 0.05 and 0.1 cm misalignment.*

## Discussion

In this work, we have experimentally demonstrated the feasibility of the universal and dynamic RF to generate Bragg peaks with different broadening as desired for the treatment. As shown in Figure 4, when we align the two RF in a way that the peak of M-RF matches the peak of the N-RF, we get a maximum broadening of 2.5 cm in the Bragg peak. However, when we align the peak of M-RF to the valley of N-RF, we get a shift in the Bragg peak without a strong change in Bragg peak size. Additionally, based on the alignment of two RFs between these two extreme cases, we can achieve different broadening in the Bragg peak. The RF unit is dynamic and can be adapted for different energy layers and/or for spots near to edges of the tumor to achieve the optimal broadening while sparing healthy organs. This is shown in the example of Figure 6(c), where we dynamically changed the position between peak-to-peak and peak-to-valley alignment for the distal energy layers, to improve the distal fall-off.

As shown in Figure 6, by using a peak-to-peak configuration, we can reduce the number of energies required to generate a uniform 12 cm SOBP by 70% compared to the reference SOBP generated with pristine Bragg peaks, with slightly worse distal fall-off. The 'hybrid' configuration, mixing peak-to-peak and peak-to-valley, improves the distal fall-off with only a negligible increase in the number of energy layers (in the example, only 1 energy layer). Therefore, with our dynamic RF, it is possible to reduce the number of energy layers by more than a factor 3 without compromising the distal fall-off. Additionally, we found no compromise in neither flatness nor lateral penumbra.

To see the effect on dose delivery time with RF (peak-to-peak configuration), we irradiated different geometrical shapes at our facility with Gantry 2. As shown in Figure 7, with the use of RF we managed to reduce the dose delivery time by 50% with an energy switching time of approximately 100 ms. This reduction in delivery time will be facility dependent, as it will depend on planning strategy and energy switching times; however, for most facilities worldwide, our RF design may result in a dose delivery time reduction by 50 to 80%, considering an energy change time between 300 ms to 2000 ms. Moreover, a synchrotron produces sharper Bragg peaks than cyclotrons; particularly for shallow targets, that usually require many energies, our RF design could bring significant advantages for synchrotron-based facilities with all types of particles, and could reduce treatment delivery time by 80-90% compared to conventional delivery without a RF [40].

As a part of the feasibility study, we also performed an extensive robustness analysis. Our RF design works perfectly with parallel beams (parallel scanning) and it is independent of spot position. The main limitation comes with downstream scanning: in this case, our RF works only for a small scanning area of 16 cm *16 cm. Additional challenges to practical realization in the clinic is the high alignment accuracy required (greater than 0.5 mm), and the speed of the motor drive needed to use the dynamic nature of the RF without adding dead time.

Our treatment delivery technique with RF is currently at the experimental stage; for clinical translation, further intensive investigation of treatment planning with different tumors is necessary. We are currently investigating an implementation in our in-house treatment planning system of plan optimization with different configurations of the RF (including the 'hybrid' one).

In our proof of principle study we have shown experimentally a RF design that adapts the Bragg peak broadening for a spot and/or energy layer based on the requirement of the dose shaping. Additionally, our RF could be already used for small animal irradiation studies with FLASH dose rates, as the peak-to-peak configuration of our RF gives 2.5 cm SOBP for low energy beams. In a longer term, this innovation could enable ultra-fast treatment delivery, with positive impact on efficiency when using motion mitigation techniques, clinical implementation of FLASH irradiations, patient throughput, and costs of treatment.

## Conclusion

In this article, a concept for a dynamic and universal RF has been introduced and its ability to achieve variable Bragg peak broadening demonstrated using both Monte Carlo simulations and measurements. With the design described here, maximum broadenings of the Bragg peak up to 2.5 cm could be achieved and the number of energy layers required to generate different SOBP could be reduced by a factor of three independently of maximum energy, corresponding to a reduction in delivery time of 50% without compromising flatness or penumbra. As this design is not patient-specific or beam model-specific, it is easily adaptable for other particle therapy facilities (both, cyclotron and synchrotron-based facilities) too.

In addition, by moving the RF a few millimeters, the new design provides flexibility in selecting different Bragg peak broadenings for a spot and/or energy layer based on the requirement of the target shape without making any changes in beamline or nozzle design. As such, it could potentially ease the treatment of mobile tumors using currently time consuming techniques such as gating and breath-hold.